\documentclass[]{aastex}

\RequirePackage{color}

\newcommand{\emaila}{jmould@swin.edu.au}

\begin{document}
\title{Modified Gravity and Large Scale Flows}
\shorttitle{Large Scale Flows}
\shortauthors{Mould et al.}
 
\author{Jeremy Mould$^{1,8}$, Matthew Colless$^{2}$, %Tamara Davis$^{10,8}$, 
Pirin Erdo{\u g}du$^{3}$, Heath Jones$^{4}$, 
John Lucey$^{5}$, Yin-Zhe Ma$^{9}$, Christina Magoulas$^{6}$ \& Christopher M. Springob$^{7,8}$}
\altaffiltext{1}{ Centre for Astrophysics \& Supercomputing, Swinburne University of Technology, Victoria 3122, Australia}
\altaffiltext{2}{Research School of Astronomy \& Astrophysics, Australian National University}
\altaffiltext{3}{American University of Kuwait}
\altaffiltext{4}{Macquarie University}
\altaffiltext{5}{Durham University}
\altaffiltext{6}{University of Cape Town}%, South Africa} 
\altaffiltext{7}{University of Western Australia/ICRAR}
\altaffiltext{8}{CAASTRO \& AAO}
\altaffiltext{9}{Jodrell Bank Centre for Astrophysics, University of Manchester}
%\altaffiltext{10}{University of Queensland}

\email{\emaila}

\begin{abstract}

Reconstruction of the local velocity field from the overdensity field and a gravitational acceleration that falls off from a point mass as r$^{-2}$ yields velocities in broad agreement with peculiar velocities measured with galaxy distance indicators.  MONDian gravity does not. To quantify this,
we introduce the velocity angular correlation function as a diagnostic of peculiar velocity field alignment and coherence as a function of scale. It is independent of the bias parameter of structure formation in the standard model of cosmology and the acceleration parameter of MOND. A modified gravity acceleration consistent with observed large scale structure would need to asymptote to zero at large distances more like  r$^{-2}$, than r$^{-1}$.
\end{abstract}


\section{Introduction}
One of the most significant products of redshift surveys is a map of large scale structure. This in turn allows us to calculate the velocity field induced by density contrasts over cosmic time. For this we often use the linear approximation that the acceleration of a galaxy does not change much over time and that velocities are not just dimensionally equivalent to acceleration multiplied by the Hubble time, but also proportional to it. Regions of high overdensity are to be avoided when using the linear approximation, as turnaround and virialization follow the rise of galaxy density to high levels.
In the era of precision cosmology, when measuring the Hubble Constant to 1\% is our aspiration (Bennett et al 2014, Suyu et al 2012) for a variety of compelling physical reasons, peculiar velocities need to be better measured and calculated by local redshift surveys. The state of the art is illustrated by Lavaux \& Tully (2010) and Magoulas et al (2012).
The fact that approximately 70\% of the Universe is dark energy and that dark energy is not physically understood (Bin\'{e}truy 2013) suggests that we should not ignore alternatives to Newton's gravity and Einstein's gravity at scales larger than those of classical GR tests. Modified gravity laws cannot yet be ruled out. In this paper we explore one such gravity law applied to the 2MRS density distribution
(Huchra et al 2012), namely Milgrom's Modification of Newtonian Gravity (MOND) (Milgrom 1983). We find that, while well motivated for kpc scales, it predicts a velocity field different from what we have observed, for example in the 6dF Galaxy Survey (Jones et al 2009).

%This investigation is clearly limited in scope to the simplest MOND referred to by Milgrom (2010) as pristine MOND. 
The unification of MOND with space expanding on Mpc scales with a scale factor $a$ is a work in progress. Close to 40 years' history of MOND has been reviewed by Sanders (2015) and Bothun (2015).
There is a general problem with all attempts to address large scale structure
problems within the MONDian framework: the framework does not exist!
There is no cosmological MOND theory is  the standard answer of MONDian aficionados. Be that as it may,
the growth of density inhomogeneities ($\delta \rho/\rho$) in that theory has been studied
by Nusser (2002) and Llinares (2008, 2014).
We note that %between the fundamental equation for the growth of structure from a mean density $\bar{\rho}$ (Peacock 1999)
the peculiar acceleration due to an overdensity in Newtonian gravity is given by Peacock (1999)

$$\dot{\bf u} + 2 \frac{\dot{a}}{a} {\bf u} = - \frac{\bf g} {a} \eqno(1n)$$

\noindent where peculiar velocity {\bf v} = a {\bf u} and %equation (2) below, the conventional Poisson equation is applied. Multifield developments of MOND adopt a modified Poisson equation. 
by Nusser (2002) as

$$\dot{\bf u} + 2 \frac{\dot{a}}{a} {\bf u} = - \surd \frac{3\Omega_m H^2 g_0} {2a} \frac{\bf g_N}{\surd g_N}\eqno(1m)$$

\noindent in the curl-free MONDian case with $\Omega_m$ the present matter density. To avoid ambiguity we write the MOND acceleration parameter a$_0$ as $g_0$.
Our purpose in this paper is not to join this development of MOND or TeVeS (Bekenstein 2004) to deal with groups of galaxies or cosmological simulations (Angus et al 2013); rather we wish to motivate the extension of peculiar velocity surveys beyond 6dFGS by illustrating the power of peculiar velocities to investigate both structure and gravity on the largest scales.

\section{Implementation}
For calculating peculiar velocities from 2MRS we have followed Erdo{\u g}du et al (2006) and used the formulation by Peebles (1980) and Davis et al (2011) in the usual notation with {\bf g(r)} representing the gravitational acceleration at {\bf r}

$$ \bf{v(r)} = \frac{{\rm 2\Omega^{4/7}\beta} \bf{g(r)}}{\rm{3H_0\Omega_m}}\eqno (2) $$
where
$$\bf{g(r)} = {\rm G\bar{\rho}}\int {\rm dr^{\prime 3}} \frac{\delta\rho^\prime}{\rho^\prime}\frac{\bf{r-r^\prime}}{\bf{|r-r^\prime|}^3} \eqno(3)$$
and $\beta~=~\Omega_m^{4/7}/b$ and $b$ the bias parameter.
On Mpc scales we are in the `deep-MOND' regime
(Zhao et al 2013) beyond the interpolation formulae between MOND
and Newtonian gravity used in the internal dynamics of galaxies, so that
the gravitational acceleration under MOND can be written as
$$ g_{\rm{MOND}} = \surd g_N \surd g_1 \eqno(4)$$ where $g_N$ is a Newtonian r$^{-2}$ acceleration field and $\surd g_1$ = $\frac{4}{3}(\surd 2 - 1) \surd g_0 $ with $g_0~\sim $  10$^{-10}$ m/s$^2$. Equation (1) of Zhao et al with $y~>>~1$
yields this definition of $g_1$.

Our calculation therefore proceeds by substituting the MONDian acceleration for the Newtonian one in equation (2). The value of $\beta$ is calculated from
the bias factor, measured for this sample to be $b$ = 1.48 $\pm$ 0.27 (Beutler et al 2012). Nusser (2014) has pointed out that not only are we assuming the linear approximation in doing this, and thus
erring in high density regions, but also we are neglecting velocities generated at early times\footnote{During early cosmic times, all accelerations on all scales are large, so that 
the Newtonian equations pertain. As time goes by, the gravitational field decreases in amplitude and 
enters the MOND region.  %So the extension of the standard relation is roughly 
This modification would be $T_N~ g_N ~+~ T_{MOND}~ g_{MOND}$,
where $T_N$ is the time spent in the Newtonian regime and $T_{MOND}$ is time spent in the MOND regime. 
$T_N/T_{MOND}$ depends on the amplitude of the initial fluctuations.}. Such initial peculiar velocities are subject to adiabatic decay, however, over the age of the Universe (Davis et al 2011).

\section{Results}
In this calculation, and generally in n-body codes, each particle communicates with every other particle. In the MONDian case every grid point that looks at the Shapley supercluster, sees an overdensity not fully attenuated , as the luminosity field is, by r$^{-2}$ and wants to move towards Shapley. The outcome of this is Figures 1 and 2 , which depict the velocity field. In Figure 1 we see a smooth flow with a coherence length as large as the volume. %In Figure 2 we see the velocity distribution function in the SGX coordinate. Figure 2 is not the problem. There are two free parameters $\beta$ and, to some degree, $g_0$ that can be adjusted to bring the speed everywhere into the observer's range in the cosmic microwave background rest frame. 
It is quite unlike the observed velocity field, and there is no free parameter to remedy it. 

\begin{figure}[h]
\begin{minipage}[b]{0.5\linewidth}
\centering
\includegraphics[clip, angle=-90, width=1.35\textwidth]{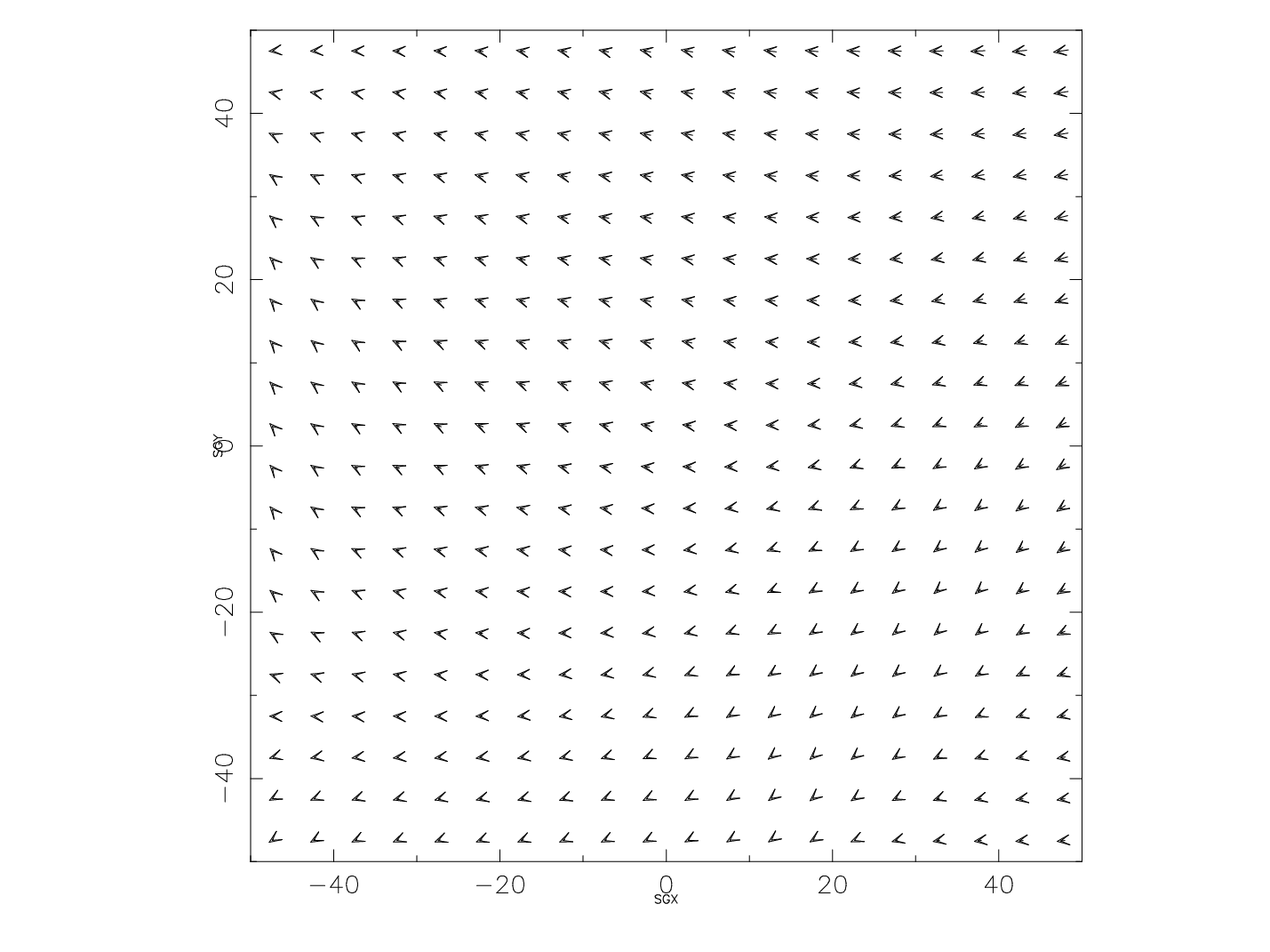}
\caption{The MONDian flowfield in the supergalactic plane. ~~~~~~~~~~~~~~~~~~~~~~~~~~~~~~~~~~~~~~~~~~~~~~~~~~~~~~~~~~~~~~~~~~~~~~~~~~~~~~~~~~~~~~~The SGX and SGY
coordinates are in units of Mpc/$h$.}
\end{minipage}
\hspace{.5 cm}
\begin{minipage}[b]{0.5\linewidth}
\centering
\includegraphics[clip, angle=-90, width=1.35\textwidth]{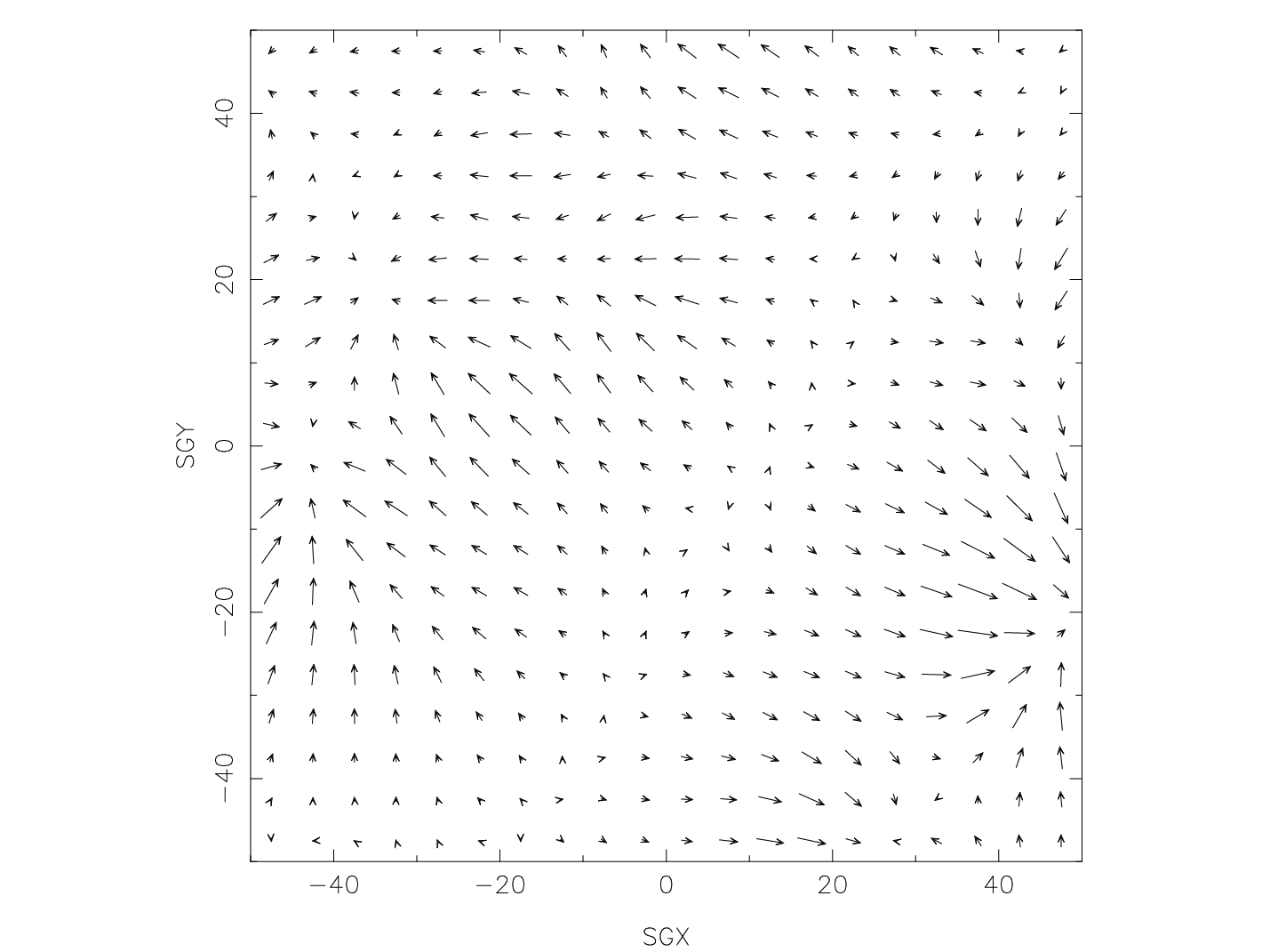}
%{/nfs/cluster/mould/fp/pirinsmall.ps}
\caption{The Newtonian flowfield for comparison with Figure 1. Prominent features are the Great Attractor on the left and the Perseus-Pisces supercluster on the right.}
\end{minipage}
\end{figure}

\begin{figure}[h]
\begin{minipage}[b]{0.5\linewidth}
\centering
\includegraphics[clip, angle=-90, width=1.0\textwidth]{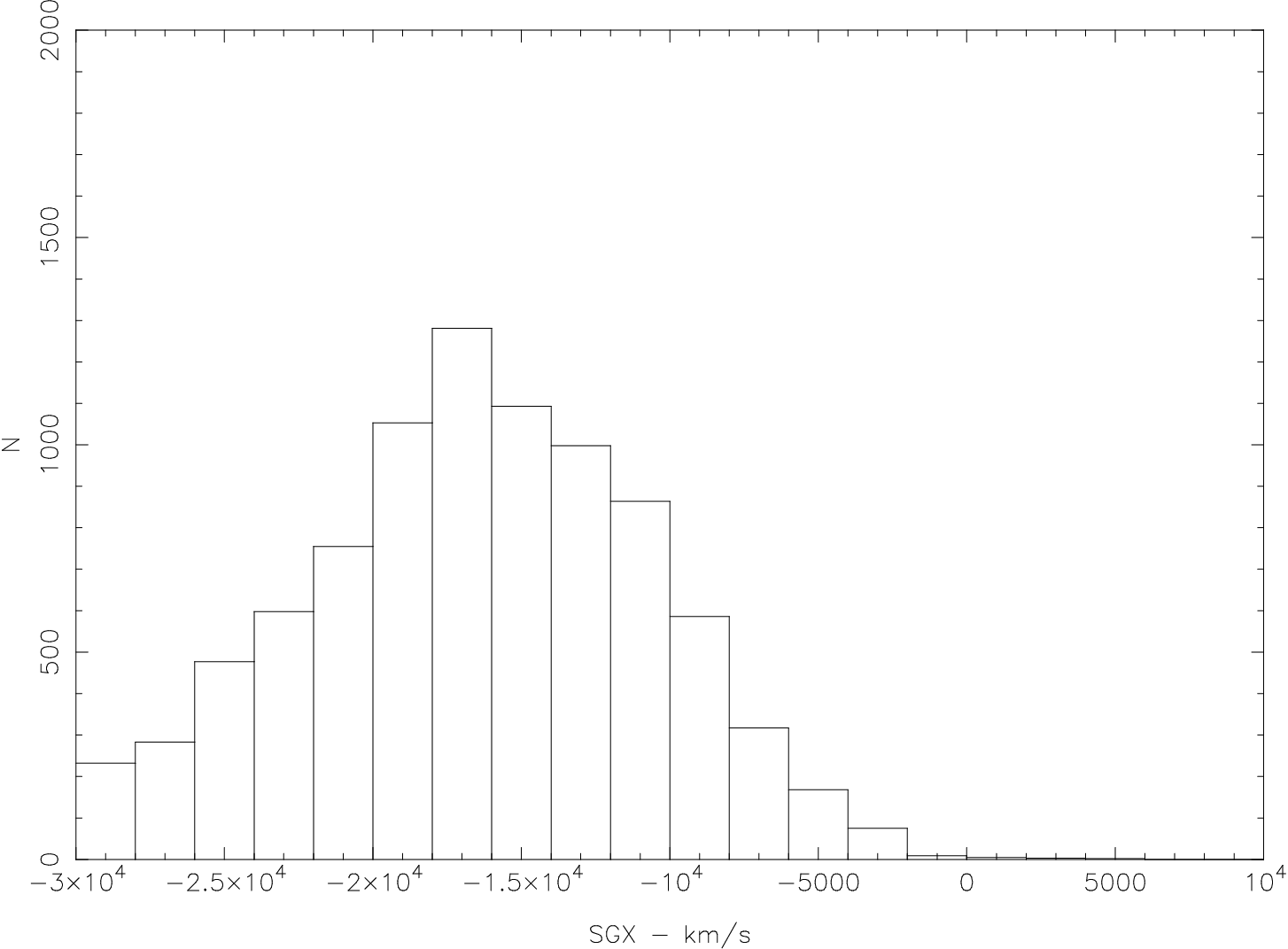}
\caption{The distribution of MONDian peculiar velocities in the SGX direction. This figure is for $\beta$ = 0.4 in equation (2), and velocities would scale
by a factor of 1.5 for $\beta$ = 0.6.}
%\vskip -0.15 truein
\end{minipage}
\hspace{.5 cm}
\begin{minipage}[b]{0.5\linewidth}
\centering
\includegraphics[clip, angle=-90, width=1.0\textwidth]{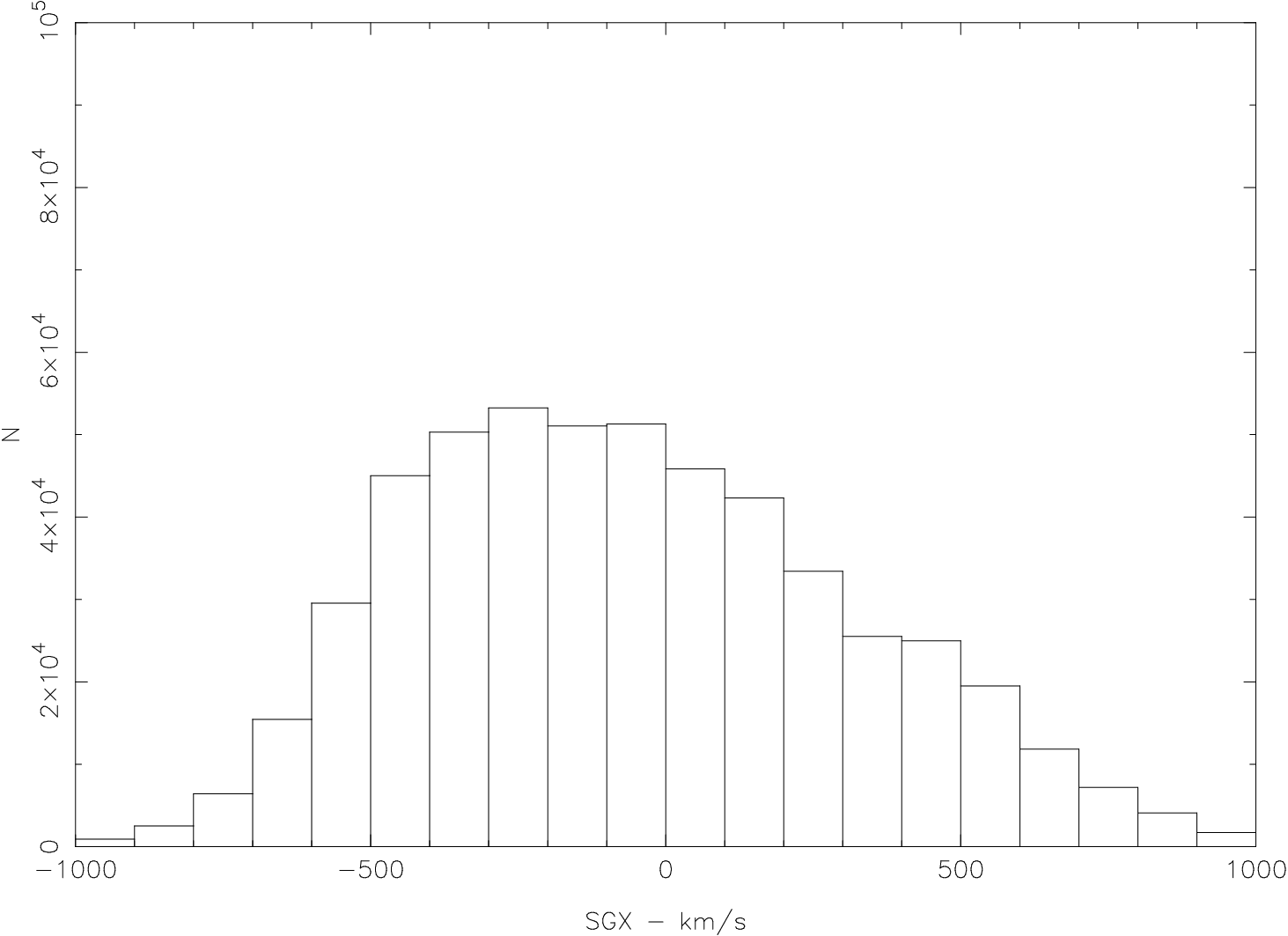}
\caption{The distribution of Erdo{\u g}du model peculiar velocities. Again, we used $\beta$ = 0.4.}
%\vskip -0.15 truein
\end{minipage}
\end{figure}

In Figures 3 \& 4 we see the predicted velocity distribution functions in the SGX coordinate. These figures  are not the problem. There are two free parameters $\beta$ and, to some degree, $g_0$ that can be adjusted to bring the speed everywhere into the range that we observers see in the cosmic microwave background rest frame. 
By contrast, Figures 2 and 4~%4 and 5%3 and 4 
for the Newtonian case do resemble the observed velocity field , and can be brought into agreement with it with $\beta~\approx$ 0.6 (Magoulas et al 2012, 2015).
Figure 5 shows both the density and velocity fields for standard model cosmological parameters. 

\begin{figure}
\vspace*{ -1 truein}
\hspace*{ -1 truein}
\includegraphics[scale=1., angle=-90]{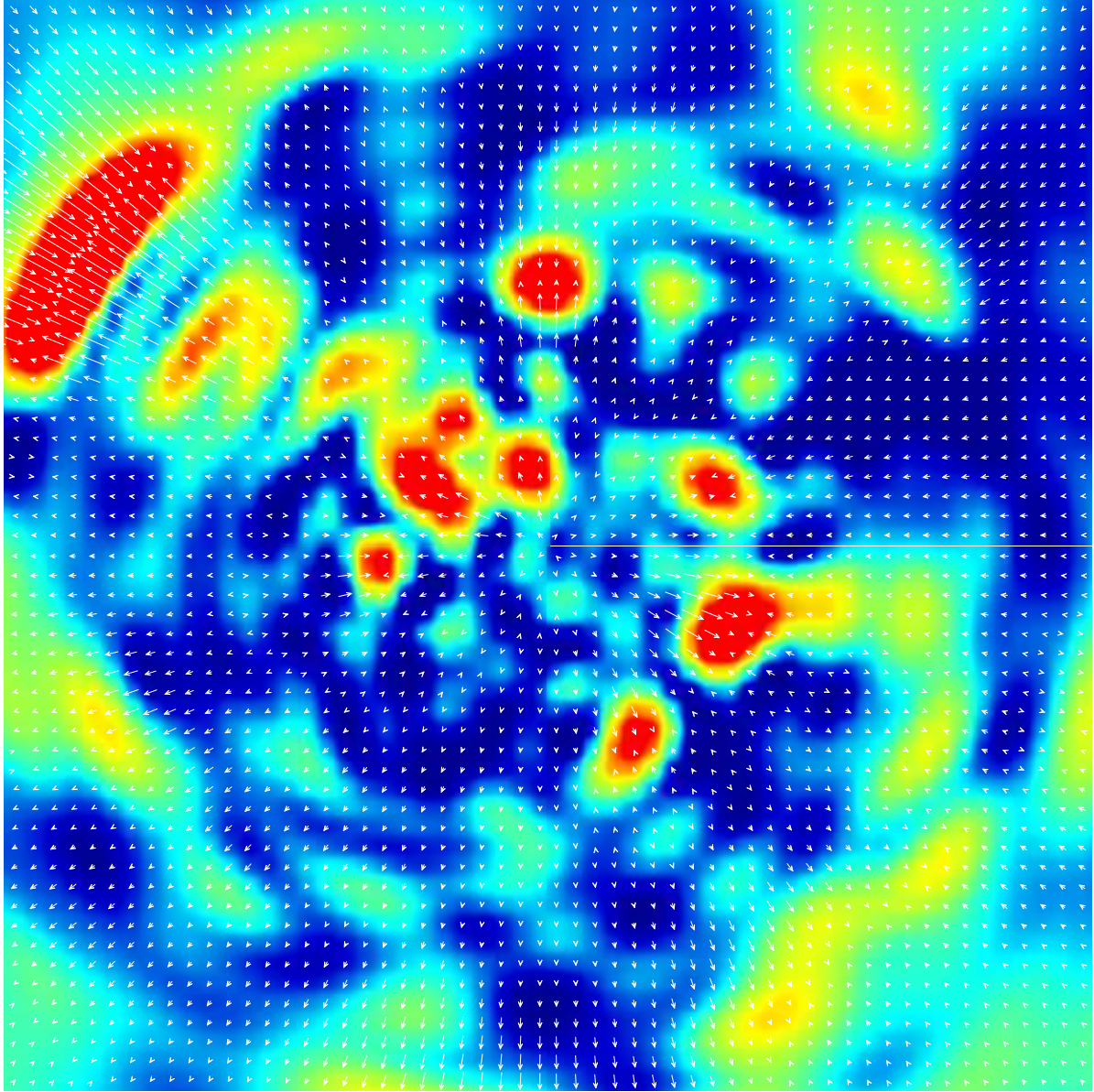}
\caption{The flow field in the supergalactic plane in the Newtonian case, superposed on the density field from 2MRS colour coded 
(red being denser than the mean by a factor of 12 and blue zero density).
We are at the origin and the two closest prominent features are inflow into the Great Attractor (towards the upper left)
and into Perseus Pisces (towards the lower right).
The longest arrows reach 1500 km/s.}

\end{figure}

\section{Analysis}
\begin{figure}
\includegraphics[scale=0.65, angle=-90]{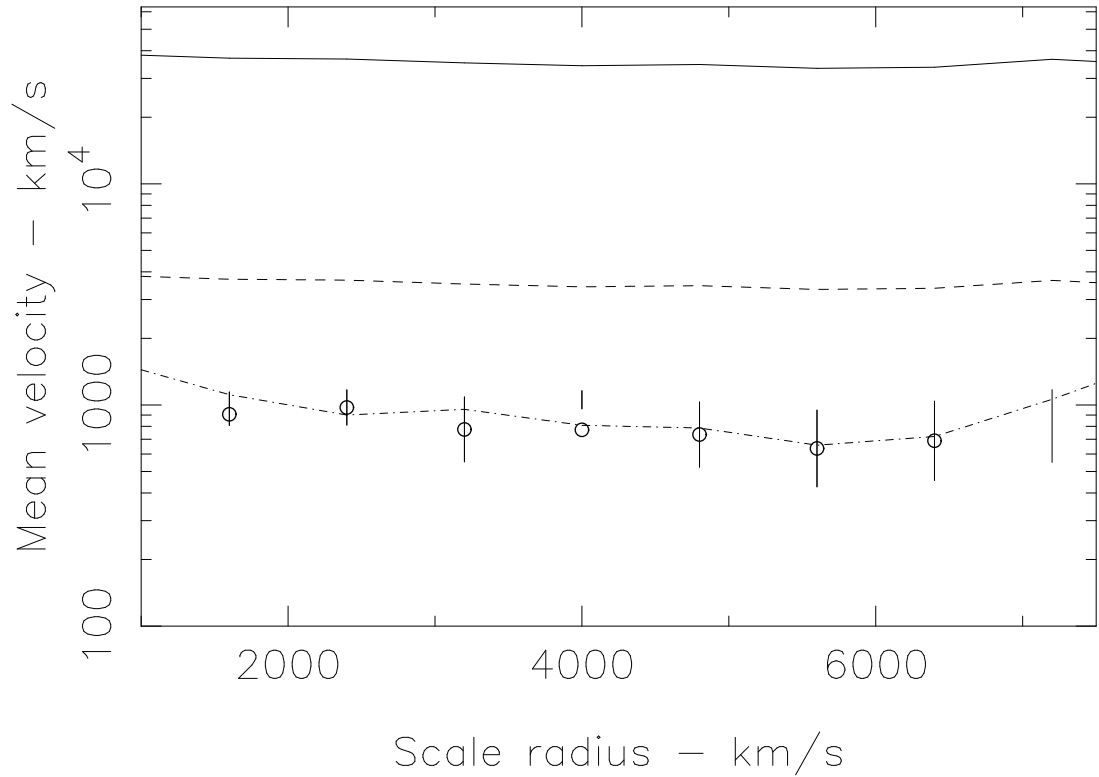}%vpowermond.ps}
\caption{Bulk flow velocity for the MOND case for $\beta$ = 0.4. 
Reducing g$_0$ by two orders of magnitude gives the dashed line. It
improves the MOND prediction but is still far from a fit to 6dFGS data:
open circles with error bars. The 2MRS model predictions for $\beta$ = 0.6
are the dot-dashed line.}
\end{figure} 

A formal comparison of the prediction of MOND and 6dFGS observations is made
in Figure 6. %5. 
Here we show bulk flow velocity  as a function of
scale. To calculate this, we create a large number of spheres of particular radii
and average the velocities within each. To reconcile
MOND and observations in this plot would require a four order of magnitude change in $g_0$, which would disrupt the agreement between MOND predictions
and galaxy rotation curves (Swaters et al 2010). The mismatch 
between our observations and MOND rules out MOND. As we see below, the standard Erdo{\u g}du r$^{-2}$
model, on the other hand, agrees with the observations within the uncertainties.

\begin{figure}
\includegraphics[scale=0.65, angle=-90]{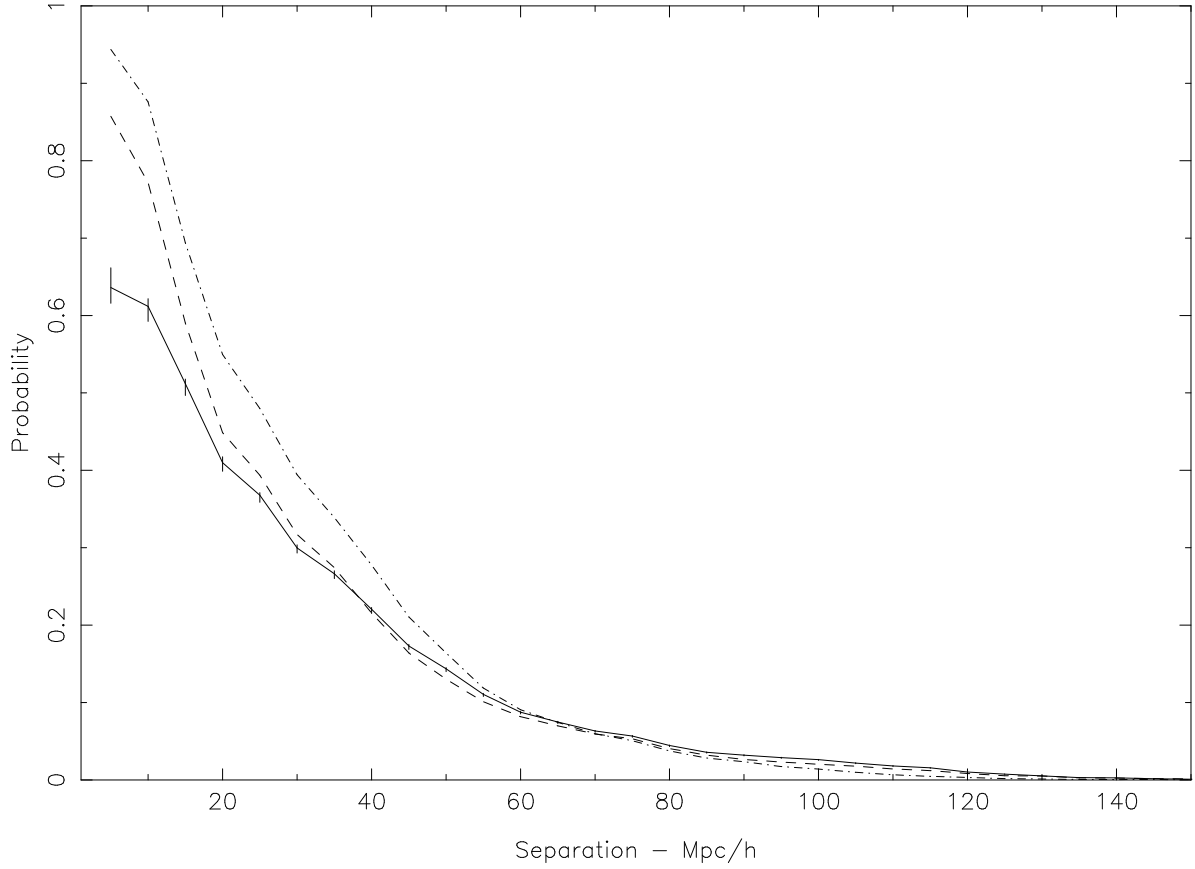}
\caption{The velocity angular correlation function for the MOND case (dot-dashed line) for the Newtonian model (dashed line)
and for 6dFGS (solid line with error bars). In the 2MRS %is denominated in redshift space and so 
galaxy separations are measured in Mpc/h,
where $h$ is the Hubble constant in units of 100 km/s.}
\end{figure} 

We have also calculated the velocity angular correlation function
as follows. For every pair of galaxies in the 6dFGS peculiar velocity sample
the angle between the radial peculiar velocities is calculated. Figure 7 %6
shows the probability that this angle $\theta$ is small ($\cos \theta~>$ 0.9)
as a function of separation.  In MOND small misalignments continue to large
galaxy separations. In the Erdo{\u g}du model the fall off is more rapid.
Again, the data are most inconsistent with MOND. For galaxy separations between
20/h and 100/h Mpc $\chi^2$ per degree of freedom is over five times larger
for MOND than it is for the $r^{-2}$ prediction. Absolute values of $\chi^2$
are hard to calculate exactly because of the expected failure of the linear approximation
at separations smaller than 20 Mpc and the non-gaussian probability distributions of 6dFGS
peculiar velocities (Springob et al 2014).
The coherence length of velocity structure measured as an e-folding scale for
this function is 2600 km/s for 6dFGS, 2700 km/s for the Newtonian 2MRS model and 
3300 km/sec for the MOND model.

\section{Conclusions}
\noindent Peculiar velocities are not a unique probe of modified gravity at the 10 Mpc scale. Weak lensing coupled with galaxy redshifts also provides a good constraint (Reyes et al 2010). Focussing on peculiar velocities, however, we conclude\\
(1) MOND predicts a velocity field overwhelmingly dominated by the largest overdensities on the largest scales (100 Mpc) that we have tested here.
The velocity angular correlation function shows markedly worse agreement
with 6dFGS in the MONDian case than in the acceleration $\sim~r^{-2}$ case.\\
(2) Smaller well established 
features observed in the flow field such as the infall into the Great Attractor (e.g. Lynden Bell et al 1987,
Mathewson \& Ford 1994) and into the Perseus Pisces supercluster (e.g. Han \& Mould 1992) are not seen in the MOND flow field.\\
(3) If we consider modified gravities more broadly than MOND, those with accelerations that fall off more slowly than r$^{-2}$ will tend to run into similar problems, but these would need to be statistically tested for a mismatch with peculiar velocity data.\\
(4) The velocity power spectrum (Johnson et al 2014) is a fine basis for such tests. Evidence for more power on large scales than $\Lambda$CDM predicts under the linear approximation and standard gravity is at the 2$\sigma$ level currently (Feldman et al 2011). Larger scale coherence than discussed here is seen (Tully 1989, Tully et al 2014). The relevance of modified gravity to such observations remains to be seen.

\acknowledgements
%\noindent Acknowledgements\\
This research was supported by the Munich Institute for Astro- and Particle Physics (MIAPP) of the DFG cluster of excellence ``Origin and Structure of the Universe".
We thank MIAPP for their hospitality while this contribution was being prepared and Richard Anderson for suggesting it. The 6dFGS is a project of the Australian Astronomical Observatory. We are grateful for grant LP130100286 from the Australian Research Council and to CAASTRO\footnote{http://www.caastro.org} for conference travel support. CAASTRO is the ARC's Centre of Excellence for All Sky Astrophysics. Comments
from David Parkinson in connection with the 2nd CAASTRO/CoEPP joint workshop on dark matter are appreciated, as were comments from the referee.

\section*{References}
\noindent  Angus, G  et al 2013, MNRAS, 436, 202\\	
Beckenstein, J 2004, Phys Rev D {\bf 70} 083509\\
Bennett, C et al 2014, ApJ, 794, 135\\
Beutler, F et al 2012, MNRAS, 423, 3420\\
Bin\'{e}truy, P 2013, AAR, 21, 67\\
Bothun, G 2015, Can J Phys, 93, 139\\
Erdo{\u g}du, P et al 2006, MNRAS, 373, 45\\
Davis, M et al 2011, MNRAS, 413, 2906\\
Feldman, H et al 2010, MNRAS, 407, 2328\\
Han, M \& Mould, J 1992, ApJ, 396, 453\\
Huchra, J et al 2012, ApJS, 199, 26\\	
Johnson, A et al 2014, MNRAS, 444, 3926 \\
Jones, DH et al 2009, MNRAS, 399, 683\\
Lavaux, G \& Tully, RB 2010, ApJ, 709, 483\\
Llinares, C 2008, MNRAS, 391, 1778\\
Llinares, C 2014, PhRvD, 89, 4023\\
Lynden-Bell, D et al 1988, ApJ, 326, 19\\
Magoulas, C et al 2012, MNRAS, 427, 245\\
Magoulas, C et al 2015, in preparation\\
Mathewson, D \& Ford, V 1994, ApJ, 434, L39\\
Milgrom, M 1983, ApJ, 270, 365\\
%Milgrom, M 2010, MNRAS, 403, 886\\
Nusser, A 2002, MNRAS, 331, 909\\
Nusser, A 2014, private communication\\
Peacock, J 1999, {\it Cosmological Physics}, Cambridge University Press\\
Peebles, P 1980, {\it The Large Scale Structure of the Universe}, Princeton University Press\\
Reyes, R et al 2010, Nature, 464, 256\\
Sanders, R 2015, Can J Phys, 93, 126\\
Springob, C et al 2014, MNRAS, 445, 2677\\
Suyu, S et al 2012, astro-ph 1202.4459\\
Swaters, R et al 2010, ApJ, 718, 380\\
Tully, RB 1989, ASSL, 151, 41 \\
Tully, RB et al 2014, Nature, 513, 71\\
Zhao, H et al 2013, A\&A, 557, L3

\end{document}